\pdfoutput=1
\documentclass[a4paper,USenglish,cleveref, autoref, thm-restate,authorcolumns]{lipics-v2019}

\title{Succinct Filters for Sets of Unknown Sizes} 

\author{Mingmou Liu}{State Key Laboratory for Novel Software Technology, Nanjing University, China}{liu.mingmou@smail.nju.edu.cn}{}{Part of the research was done when Mingmou Liu was visiting the Princeton University.}

\author{Yitong Yin}{State Key Laboratory for Novel Software Technology, Nanjing University, China}{yinyt@nju.edu.cn}{}{}

\author{Huacheng Yu}{Princeton University, Princeton, New Jersey, United States}{yuhch123@gmail.com}{}{}

\authorrunning{M.~Liu, Y.~Yin, and H.~Yu} 

\Copyright{Mingmou Liu, Yitong Yin, and Huacheng Yu} 



\funding{Mingmou Liu and Yitong Yin are supported by National Key R\&D Program of China 2018YFB1003202 and NSFC under Grant Nos.~61722207 and 61672275.}

\nolinenumbers 

\hideLIPIcs  

\EventEditors{Artur Czumaj, Anuj Dawar, and Emanuela Merelli}
\EventNoEds{3}
\EventLongTitle{47th International Colloquium on Automata, Languages, and Programming (ICALP 2020)}
\EventShortTitle{ICALP 2020}
\EventAcronym{ICALP}
\EventYear{2020}
\EventDate{July 8--11, 2020}
\EventLocation{Saarbrücken, Germany (virtual conference)}
\EventLogo{}
\SeriesVolume{168}
\ArticleNo{79}

\usepackage[utf8]{inputenc}

\usepackage{hyperref}
\usepackage[boxruled,ruled]{algorithm2e}

\usepackage{amsmath}
\usepackage{amsfonts}
\usepackage{amssymb}
\usepackage{tikz}
\usepackage{mathrsfs}
\usepackage{multirow}
\usepackage[colorinlistoftodos]{todonotes}

\usepackage{amsthm}
\usepackage{cases}

\newcommand{\HammingCube}[1]{\{0,1\}^{#1}}

\newcommand{\cD}{\mathcal{D}}

\newcommand{\PFC}{\mathtt{PFC}}

\newcommand{\poly}{\mathrm{poly}}

\newcommand{\PMInit}[1]{\mathsf{initialize}({#1})}
\newcommand{\PMDestr}[1]{\mathsf{destroy}({#1})}
\newcommand{\PMInsert}[2]{\mathsf{insert}({#1},{#2})}
\newcommand{\PMQuery}[2]{\mathsf{query}({#1},{#2})}
\newcommand{\PMDecr}[1]{\mathsf{decrement}({#1})}

\def\mainfile{}
\begin{document}

\maketitle

\begin{abstract}
  The membership problem asks to maintain a set $S\subseteq[u]$, supporting insertions and \emph{membership queries}, i.e., testing if a given element is in the set.
  A data structure that computes exact answers is called a \emph{dictionary}.
  When a (small) false positive rate $\epsilon$ is allowed, the data structure is called a \emph{filter}.

  The space usages of the standard dictionaries or filters usually depend on the upper bound on the size of $S$, while the actual set can be much smaller.

  Pagh, Segev and Wieder~\cite{pagh2013approximate} were the first to study filters with varying space usage based on the \emph{current} $|S|$.
  They showed in order to match the space with the current set size $n=|S|$, any filter data structure must use $(1-o(1))n(\log(1/\epsilon)+(1-O(\epsilon))\log\log n)$ bits, in contrast to the well-known lower bound of $N\log(1/\epsilon)$ bits, where $N$ is an upper bound on $|S|$.
  They also presented a data structure with almost optimal space of $(1+o(1))n(\log(1/\epsilon)+O(\log\log n))$ bits provided that $n>u^{0.001}$, with expected amortized constant insertion time and worst-case constant lookup time.

  In this work, we present a filter data structure with improvements in two aspects:
  \begin{itemize}
      \item it has constant worst-case time for all insertions and lookups with high probability;
      \item it uses space $(1+o(1))n(\log (1/\epsilon)+\log\log n)$ bits when $n>u^{0.001}$, achieving optimal leading constant for all $\epsilon=o(1)$.
  \end{itemize}
  We also present a dictionary that uses $(1+o(1))n\log(u/n)$ bits of space, matching the optimal space in terms of the current size, and performs all operations in constant time with high probability.
\end{abstract}





\section{Introduction}
%

Membership data structures are fundamental subroutines in many applications, including databases~\cite{10.1145/1365815.1365816}, content delivery network for web caching~\cite{10.1145/2805789.2805800}, image processing~\cite{Jiang2018}, scanning for viruses~\cite{1577953}, etc.
The data structure maintains a set of keys from a key space $[u]$,\footnote{Throughout the paper, $[u]$ stands for the set $\{0,\ldots,u-1\}$.} supporting the following two basic operations:
\begin{itemize}
  \item insert($x$): insert $x$ into the set;
  \item lookup($x$): return \texttt{YES} if $x$ is in the set, and \texttt{NO} otherwise.
\end{itemize}
When false positive errors are allowed, such a data structure usually is referred as a \emph{filter}.
That is, a filter with false positive rate $\epsilon$ may answer \texttt{YES} with probability $\epsilon$ when $x$ is not in the set (but it still needs to always answer \texttt{YES} when $x$ is in the set).

In the standard implementations, a initialization procedure receives the key space size $u$ and a \emph{capacity} $N$, i.e., an upper bound on the number of keys that can simultaneously exist in the database.
Then it allocates sufficient space for the data structure, e.g., a hash table consisting of $\Theta(N)$ buckets.
Thereafter, the memory usage is always staying at the maximum, as much space as $N$ keys would take.
It introduces inefficiency in the space, when only few keys have been inserted so far.
On the other hand, it could also happen that only a rough estimation of the maximum size is known (e.g. ~\cite{guo2006theory, almeida2007scalable, doi:10.1177/1094342015618452}).
Therefore, to avoid overflowing, one has to set the capacity conservatively.
The capacity parameter given to the initialization procedure may be much more than the actual need.
To avoid such space losses, a viable approach is to dynamically allocate space such that at any time, the data structure occupies space depending only on the \emph{current} database size (rather than the maximum possible).

For exact membership data structures, it turns out that such promise is not too hard to obtain if one is willing to sacrifice an extra constant factor in space and accept amortization: When the current database has $n$ keys, we set the capacity to $2n$; after $n$ more keys are inserted, we construct a new data structure with capacity equal to $4n$ and transfer the whole database over.
The amortized cost to transfer the database is $O(1)$ per insertion.
Raman and Rao~\cite{raman2003succinct} showed that the extra constant factor in space is avoidable, they designed a \emph{succinct}\footnote{A succinct data structure uses space equal to the information theoretical minimum plus an asymptotically smaller term called \emph{redundancy}. } membership data structure using space $(1+o(1))\log\binom{u}{n}$,\footnote{All logarithms are base $2$.} where $n$ is the \emph{current} database size, supporting insertions in expected amortized constant time, and lookup queries in worst-case constant time.

For filters, the situation is more complicated.
The optimal space to store at most $N$ keys while supporting approximate membership queries with false positive rate $\epsilon$ is $N\log1/\epsilon$ ~\cite{carter1978exact, lovett2010lower} (Pagh, Pagh and Rao~\cite{pagh2005optimal} achieved $(1+o(1))N\log 1/\epsilon$ bits).
However, the above trick to reduce the space may not work in general.
This is because the filter data structures do not store perfect information about the database, and therefore, it is non-trivial to transfer to the new data structure with capacity $4n$, as one might not be able to recover the whole database from the previous data structure.
In fact, Pagh, Segev and Wieder~\cite{pagh2013approximate} showed an information theoretical space lower bound of $(1-o(1))n(\log1/\epsilon+(1-O(\epsilon))\log\log n)$ bits, regardless of the insertion and query times.
That is, one has to pay extra $\approx\log\log n$ bits per key in order to match the space with the current database size.
They also proposed a data structure with a nearly matching space of $(1+o(1))n\log1/\epsilon+O(n\log\log n)$ bits when $n>u^{0.001}$, while supporting insertions in expected amortized constant time and lookup queries in worst-case constant time.
When $\epsilon$ is at least $1/\poly\log n$, the extra $\log\log n$ bits per key is dominating.
It was proposed as an open problem in~\cite{pagh2013approximate} whether one can make the $\log\log n$ term succinct as well, i.e., to pin down its leading constant.

On the other hand, an amortized performance guarantee is highly undesirable in many applications.
For instances, IP address lookups in the context of router hardware ~\cite{broder2001using, kirsch2007using}, and timing attacks in cryptography ~\cite{lipton1993clocked, kocher1996timing, osvik2006cache, naor2019bloom}. 
When the database size is always close to the capacity (or when the space is not a concern), it was known how to support all operations in worst-case constant time~\cite{dietzfelbinger1990new,arbitman2010backyard} with high probability.
That is, except for a probability of $1/\poly\, n$, the data structure handles \emph{every} operation in a sequence of length $\poly\, n$ in constant time.\footnote{This is stronger guarantee than expected constant time, since when the unlikely event happened, one could simply rebuild the data structure in linear time. The expected time is still a constant.}
However, it was not known how to obtain such a guarantee when the space is succinct with respect to the current database size, i.e., $(1+o(1))\log\binom{u}{n}$.
For filters, Pagh et al.~\cite{pagh2013approximate} showed it is possible to get worst-case constant time with high probability, at the price of a constant factor more space $O(n\log 1/\epsilon+n\log\log n)$.
They asked if there is a data structure which enjoys the succinct space usage and the worst-case constant time with high probability simultaneously.

\subsection{Main Results}
In this paper, we design a new dynamic filter data structure that answers both questions.
Our data structure has both worst-case constant time with high probability and is succinct in space in terms of the current database size.

\begin{theorem}[Dynamic filter - informal]\label{main-thm:filter-informal}
There is a data structure for approximate membership with false positive rate $\epsilon$  that
    uses space $(1+o(1))n(\log(1/\epsilon)+\log\log n)$ bits, where $n>u^{0.001}$ is the current number of keys in the database, 
    such that every insertion and lookup takes constant time in the worst case with high probability.
\end{theorem}

We also present a dictionary data structure with the space depending on the current $n$.
A dictionary is a generalization of membership data structures, it maintains a set of key-value pairs, supporting
\begin{itemize}
	\item insert($x, y$): insert a key-value pair $(x, y)$ for $x\in [u]$ and $v$-bit $y$ ;
	\item lookup($x$): if $\exists (x, y)$ in the database, output $y$; otherwise output \texttt{NO}.
\end{itemize}
By setting $v=0$, the lookup query simply tests if $x$ is in the database.

\begin{theorem}[Dynamic dictionary - informal]\label{main-thm:dictionary-informal}
There is a dictionary data structure that
    uses space $(1+o(1))n(\log(u/n)+v+O(\log\log\log u))$ bits, where $n>u^{0.001}$ is the current number of key-value pairs in the database, 
    such that every insertion and lookup takes constant time in the worst case with high probability.
\end{theorem}

\subsection{Related Work}
\subparagraph{Membership with Constant Time Worst-Case Guarantee.}
The FKS perfect hashing~\cite{FKS84} stores a set of $n$ \emph{fixed} (i.e., static) keys using $O(n)$ space, supporting membership queries in worst-case constant time.
Dietzfelbinger, Karlin, Mehlhorn, Meyer auf der Heide, Rohnert and Tarjan~\cite{dietzfelbinger1988dynamic} introduced an extension of the FKS hashing, which is the first dynamic membership data structure with worst-case constant query time and the expected amortized constant insertion time.
Later, Dietzfelbinger and Meyer auf der Heide~\cite{dietzfelbinger1990new} improved the insertion time to worst-case constant, with an overall failure probability of $1/\poly\, n$.
Demaine, Meyer auf der Heide, Pagh and P{\v{a}}tra{\c{s}}cu~\cite{demaine2006dictionariis} improved the space to $O(n\log(u/n))$ bits of space.
Arbitman, Naor and Segev~\cite{arbitman2009amortized} proved that a de-amortized version of cuckoo hashing~\cite{kirsch2007using} has constant operation time in the worst case with high probability.

On the other hand, filters can be reduced to dictionaries with a hash function $h:[u]\to[n/\epsilon]$, and thus, all the dictionaries imply similar upper bounds for filters~\cite{carter1978exact}.

\subparagraph{Succinct Membership.}
Raman and Rao~\cite{raman2003succinct} presented the first succinct dictionary with constant time operations, while the insertion time is amortized.
Arbitman, Naor and Segev~\cite{arbitman2010backyard} refined the schema of~\cite{arbitman2009amortized}, suggested a succinct dictionary with worst case operation time with high probability.

By using the reduction from~\cite{carter1978exact} and the succinct dictionary from~\cite{raman2003succinct}, Pagh, Pagh and Rao~\cite{pagh2005optimal} provided a succinct filter with constant time, while the insertion time is amortized due to~\cite{raman2003succinct}.
Bender, Farach-Colton, Goswami, Johnson, McCauley and Singh~\cite{bender2018bloom} suggested a succinct \emph{adaptive filter}\footnote{In an adaptive filter, for a negative query $x$, the false positive event is independent of previous queries.} with constant time operation in the worst case with high probability.

\subparagraph{Membership for Sets of Unknown Sizes.}
The data structure of Raman and Rao~\cite{raman2003succinct} can be implemented such that the size of the data structure always depends on the ``current $n$''.
Pagh, Segev and Wieder~\cite{pagh2013approximate} were the first to study dynamic filters in this setting from a foundational perspective.
As we mentioned above, they proved an information-theoretical space lower bound of $(1-o(1))n(\log(1/\epsilon)+(1-O(\epsilon))\log\log n)$ bits for filter, and presented a filter data structure using $n(\log(1/\epsilon)+O(\log\log n))$ bits of space with constant operation time when $n>u^{0.001}$.
Indeed, the insertion time is expected amortized, since the succinct dictionary of Raman and Rao is applied as a black box (it was not clear if any succinct dictionary with worst-case operational time can be generalized to this setting).

Very recently, Bercea and Even~\cite{bercea2019fully} proposed a succinct membership data structure for maintaining \emph{dictionaries} and \emph{random multisets} with constant operation time.
While their data structure is originally designed for the case where an upper bound $N$ on the keys is given (and the space usage is allowed to depend on $N$), we note that it is possible to extend their solution and reduce the space to depend only on the current $n$.
However, their data structure assumes free randomness, and straightforward extension results in an additive $\Omega(n\log \log u)$ term in space.
The redundancy makes their data structure space-inefficient for filters, since the space lower bound is $(1-o(1))n(\log(1/\epsilon)+(1-O(\epsilon))\log\log n)$.


\subsection{Previous Construction}

As we mentioned earlier, for dynamic membership data structures, if we are willing to pay an extra constant factor in space, one way to match the space with the ``current'' $n$ is to set the capacity to be $2n$.
When the data structure is full after another $n$ insertions, we double the capacity, and transfer the database to the new data structure.
However, the standard way to construct an efficient filter is to hash $[u]$ to $[n/\epsilon]$ (where $\epsilon$ is the false positive rate) and store all $n$ hash values in a membership data structure, which takes $O(n\log 1/\epsilon)$ bits of space.
As we insert more keys and increase the capacity to $4n$, the range of the hash value needs to increase as well.
Unfortunately, it cannot be done, because the original keys are not stored, and we have lost the information in order to save space (this is exactly the point of a filter).
On the other hand, we could choose to keep the previous data structure(s), and only insert the future keys to the new data structure.
For each query, if it appears in any of the (at most $\log n$) data structures, we output \texttt{YES}.
By setting the false positive rate for the data structure with capacity $2^i$ to $O(\epsilon/i^2)$, the overall false positive rate is at most $\epsilon\cdot \sum_{i}O(1/i^2)\leq\epsilon$ by union bound.
The total space usage becomes roughly $n\log (\log^2 n/\epsilon)=n(\log 1/\epsilon+O(\log\log n))$.

To avoid querying all $\log n$ filters for each query, the previous solution by Pagh et al.~\cite{pagh2013approximate} uses a \emph{single} global hash function $h$ that maps $[u]$ to $\log(u/\epsilon)$-bit strings for all $\log n$ filters.
For a key $x$ in the $i$-th data structure (with capacity $2^i$), one simply takes the first $i+\log1/\epsilon+2\log i$ bits of $h(x)$ as its hash value.
Then querying the $i$-th data structure on $y$ is to check whether the $(i+\log1/\epsilon+2\log i)$-bit prefix of $h(y)$ exists.
Since all filters use the same hash function, the overall task is to check whether \emph{some} prefix of $h(y)$ appears in the database, which now consists of strings of various lengths.
Note that there are very few short strings in the database, the previous solution extends all short strings to length $\log (n/\epsilon)$ by duplicating the string and appending all possible suffixes, e.g., a string of length $\log(n/\epsilon)-c$ is duplicated into $2^c$ strings by appending all possible $c$-bit suffixes.
Then all strings are stored in one single dictionary (longer strings are stored according to their first $\log(n/\epsilon)$ bits), and the query becomes to check if the $\log (n/\epsilon)$-bit prefix of $h(y)$ is in the dictionary, which is solved by invoking Raman and Rao~\cite{raman2003succinct}.
One may verify that duplicating the short strings does not significantly increase the total space, and comparing only the $\log(n/\epsilon)$-bit prefix of a longer string does not increase the false positive rate by much.

\subsection{Our Techniques}

Our new construction follows a similar strategy, but the ``prefix matching'' problem is solved differently.
Given a collection of $2^{i-1}<n\leq 2^i$ strings of various lengths, we would like to construct a data structure such that given any query $h(y)$, we will be able to quickly decide if any prefix of $h(y)$ appears in the database.
The first observation is that the short strings in the database can be easily handled.
In fact, all strings shorter than $i$ bits can be stored in a ``truth table'' of size $2^i=O(n)$.
That is, we simply store for all $i$-bit strings, whether any of its prefix appears in the database.
For a query $h(y)$, by checking the corresponding entry of its $i$-bit prefix, one immediately resolves all short strings.
On the other hand, for strings longer than $\log m$ bits, we propose a new (exact) membership data structure, and show that it in fact, automatically solves prefix matching when all strings are long.
Before describing its high-level construction in Section~\ref{sec:intro_mem_data_str}, let us first see what it can do and how it is applied to our filter construction.

When the capacity is set to $m$, the membership data structure stores $n\leq m$ keys from $[u]$ using space $n(\log(u/m)+O(\log\log\log u))+O(m)$ bits, supporting insertion and membership query in worst-case constant time with high probability.
When applying to prefix matching, it stores $n$ strings of length at most $\ell$ (and more than $\log m$) using $n(\log (2^\ell/m)+O(\log\log \ell))+O(m)$ bits.
Using this data structure with the capacity set to $m=2^i$, we are able to store the database succinctly when $m/2<n\leq m$.
As we insert more keys to the database, the capacity needs to increase.
Another advantage of our membership data structure is that the data can be transferred from the old data structure with capacity $m$ to a new one with capacity $2m$ in $O(m)$ time.
More importantly, the transfer algorithm runs almost ``in-place'', and the data structure remains ``queryable'' in the middle of the execution.
That is, one does not need to keep both data structures in full, at any time the total memory usage is still $n(\log (2^\ell/n)+O(\log\log \ell))+O(m)$, and the data structure can be queried.
Therefore, as $n$ is increasing from $m/2$ to $m$, we gradually build a new data structure with capacity $2m$.
Every time a key is inserted, the background data-transfer algorithm is run for constant steps.
By the time $n$ reaches $m$, we will have already transferred everything to the new data structure, and will be ready to build the next one with capacity $4m$.
Overall, the data structure is going to have $\log n$ stages, the $i$-th stage handles the $(2^{i-1}+1)$-th to the $2^i$-th insertion.
In each stage, the database size is doubled, and the data structure also gradually doubles its capacity.
This guarantees that the total space is succinct with respect to the current database size, and every operation is handled in constant time with high probability.

Finally, to pin down the leading constant in the extra $O(\log\log n)$ bits, we show that for the $n$-th inserted key $x$ for $2^{i-1}<n\leq 2^i$, storing the $(i+\log (i/\epsilon)+\log\log\log u)$-bit prefix of $h(x)$ balances the false positive rate and the space.
Since our new membership data structure only introduces an extra $\approx \log\log i\approx\log\log\log n$ bits of space per key, it is not hard to verify that the total space of our construction is $(1+o(1))n(\log (1/\epsilon)+\log\log n)$.

\subsubsection{Membership Data Structure}\label{sec:intro_mem_data_str}
In the following, let us briefly describe how our new membership data structure works.
The data structure works in the \emph{extendable array} model, as the previous solution by Raman and Rao.
See Section~\ref{section: memory models} or~\cite{raman2003succinct} for more details.

Our main technique contribution is the idea of \emph{data block}.
Without the data blocks, our data structure degenerates into a variant of the one proposed in \cite{bercea2019fully}.
Instead of a redundancy of $O(n\log\log\log u)$ bits, the degeneration contributes a redundancy of $O(n\log\log u)$ bits, which makes the data structure space-inefficienct for filters as we discussed early.

For simplicity, let us for now assume that we have free randomness, and the first step is to randomly permute the universe.
Thus, we may assume that at any time, the database is a uniformly random set (of certain size).
We divide the universe into $m/\log u$ \emph{buckets}, e.g., according to the top $\log (m/\log u)$ bits of the key.
Then with high probability, every bucket will have $O(\log u)$ keys.
We will then dynamic allocate space for each bucket.
Note that given that a key is a bucket $b$, we automatically know that its top $\log (m/\log u)$ bits is ``$b$''.
Therefore, within each bucket, we may view the keys have lengths only $\log u-\log (m/\log u)=\log ((u\log u)/m)$, or equivalently, the universe size being $(u\log u)/m$ (recall that the goal is to store each key using $\approx\log (u/m)$ bits on average).

To store the keys in a bucket, we further divide it into \emph{data blocks} consisting of $O(\log u/\log \log u)$ keys each, based on the time of insertion.
That is, the first $O(\log u/\log \log u)$ keys inserted to this bucket will form the first data block, the next $O(\log u/\log \log u)$ keys will be the second data block, etc.
Since each data block has few enough keys, they can be stored using a \emph{static} constructions (supporting only queries) using nearly optimal space of $\approx\log \binom{(u\log u)/m}{O(\log u/\log \log u)}$, which is $\log ((u\log \log u)/m)=\log (u/m)+\log\log\log u$ bits per key, or a dynamic constructions use $\log (u/m)+O(\log \log u)$ bits per key.
The latest data block, which we always insert the new key into, is maintained using the dynamic construction. 
When it becomes full, we allocate a new data block, and at the same time, we run a in-place \emph{reorganization} algorithm in the background.
The reorganization algorithm runs in $O(\log u/\log \log u)$ time, and convert the dynamic construction into the static construction, which uses less space.
For each insertion in the future, the reorganization algorithm is run for constant steps, thus, it finishes before the next data block becomes full.
Finally, for each bucket, we maintain an \emph{adaptive prefixes} structure ~\cite{bender2017bloom, bender2018bloom} to navigate the query to the relevant data block.
Roughly speaking, when all $O(\log u)$ keys in the bucket are random, most keys will have a unique prefix of length $\log \log u$.
In fact, Bender et al.~\cite{bender2017bloom, bender2018bloom} showed that for every keys, the shortest prefix that is unique in the bucket can be implicitly maintained in constant time, and the total space for all $O(\log u)$ keys is $O(\log u)$ bits with high probability.\footnote{The $O(\log u)$-bit representation is implicit.}
We further store for each such unique prefix, which data block contains the corresponding key.
It costs $O(\log\log \log u)$ bits per key.
Given a query, the adaptive prefix structure is able to locate the prefix that matches the query in constant time, which navigates the query algorithm to the (only) relevant data block.
We present the details in Section~\ref{section: prefix-matching}.

\ifx\mainfile\undefined
\documentclass[11pt]{article}
\input{macro}

\begin{document}
\fi

\section{Preliminaries}\label{sec:prelim}
\subsection{String Notations}\label{sec:string-notation}
Let $\HammingCube{\le \ell}\triangleq\bigcup_{0\le i\le\ell}\HammingCube{i}$ and $\HammingCube{*}\triangleq\bigcup_{i\ge 0}\HammingCube{i}$.
Given a string $x\in\{0,1\}^\ell$, we use $|x|=\ell$ to denote its length.
We denote by $a\circ b$ the concatenation of two strings $a,b\in\HammingCube{*}$.
We denote the concatenation of $k$ ones or zeros by $1^k$ or $0^k$, respectively.

For $x,y\in\HammingCube{*}$, we use $x\sqsubseteq y$ (or $y\sqsupseteq x$) to denote that $y$ is a \emph{prefix} of $x$, formally:
\begin{align}
x\sqsubseteq y  \iff x=y\circ a \text{ for some }a\in\HammingCube{*}.
\end{align}
%
Note that our notation is unconventional: we use $x\sqsubseteq y$ for $y$ prefixing $x$, to reflect that the Hamming cube identified by $x$ is contained by the Hamming cube for its prefix $y$.

For two strings $x,y$ such that $|x|\le |y|$, to compare $x$ and $y$ in lexicographical order, we compare $x\circ \perp^{|y|-|x|}$ and $y$  in lexicographical order, where $\perp$ is a special symbol which is smaller than any other symbol.

Recall that an injection (code) on strings is a \emph{prefix-free code} if no codeword is a prefix of another codeword.
\begin{claim}\label{claim: prefix-free-code}
  There is a prefix-free code $\mathtt{PFC}:\HammingCube{\le\ell}\to\HammingCube{\ell+1}$ for strings of length $\le\ell$.
\end{claim}
\begin{proof}
  Given any $x\in\HammingCube{\le\ell}$, the codeword $\PFC(x)$ is $1^{\ell-|x|}\circ 0\circ x$.
\end{proof}

\subsection{Computational Models}
\subsubsection{Random Access Machine}\label{section: RAM model}
Throughout the paper, we use $w$ to denote the word size:
each memory word is a Boolean string of $w$ bits.
We assume that the total number of memory words is at most $2^w$, and each memory word has an unique address from $[2^w]$, so that any pointer fits in one memory word.
We also assume CPU has constant number of registers of size $w$, and any datapoint fits in constant number of words (i.e.~$w=\Omega(v+\log u)$).
During each CPU clock tick, CPU may load one memory word to one of its register, write the content of some register to some memory word, or execute the basic operations on the registers.
Specifically, the basic operations include four arithmetic operations (addition, subtraction, multiplication, and division),  bitwise operations (AND, OR, NOT, XOR, shifting), and comparison.

\subsubsection{Memory Models}\label{section: memory models}
We use a memory access model known as the \emph{extendable arrays}~\cite{raman2003succinct} to model the dynamic space usage.

The extendable array is one of the most fundamental data structures in practice. It is implemented by the standard libraries of most popular programming languages, such as \texttt{std::vector} in \texttt{C++}, \texttt{ArrayList} in \texttt{java} and \texttt{list} in \texttt{python}.
\begin{definition}[Extendable arrays]
  An \emph{extendable array} of length $n$ maintains a sequence of $n$ fixed-sized elements, each assigned a unique address from $[n]$, such that the following operations are supported:
  \begin{itemize}
	\item $\mathtt{access}(i)$: access the element with address $i$;
	\item $\mathtt{grow}$: increment the length $n$, creating an arbitrary element with address $n+1$;
	\item $\mathtt{shrink}$: decrement the length $n$, remove the element with address $n$.
  \end{itemize}
  A \emph{collection of extendable arrays} supports
  \begin{itemize}
	\item $\mathtt{create}(r)$: create an empty extendable array with element of size $r$ and return its name;
	\item $\mathtt{destroy}(A)$: destroy the empty extendable array $A$;
	\item $\mathtt{access}(A,i)$, $\mathtt{grow}(A)$, $\mathtt{shrink}(A)$: apply the corresponding operations on array $A$.
  \end{itemize}
  Each of above operations takes constant time.
  The space overhead of an extendable array is $O(w)+nr$, where $w,n,r$ are the word size, the length of the array, and the element size respectively.
  Indeed, the space overhead of a collection of extendable arrays is $O(|\mathcal{A}|w)+\sum_{A\in\mathcal{A}}n_Ar_A$, where $\mathcal{A}$, $n_A$ and $r_A$ are the set of extendable arrays, the length of array $A$, and the element size of array $A$ respectively.
\end{definition}
\noindent

\bigskip

We also consider the following \emph{allocate-free} model.
\begin{definition}[Allocate and free]
  In the \emph{allocate-free} model, there are two built-in procedures:
  \begin{itemize}
	\item $\mathtt{allocate}(b)$: return a pointer to a block of $b$ consecutive memory words which is uninitialized;
	\item $\mathtt{free}(p)$: free the block of consecutive memory words which is pointed by $p$ and have been initialized to $0$s.
  \end{itemize}
  Each of above operations takes constant time.
  The total space overhead is $O(|\mathcal{A}|w)+\sum_{A\in\mathcal{A}}n_Aw$, where $\mathcal{A}$ is set of all memory blocks and $n_A$ is the length of memory block $A$.
\end{definition}
\noindent
We discuss the space usages of our data structures in allocate-free model in Section \ref{section: allocate-free}.

To avoid the pointer being too expensive in the dynamic memory models, we assume $w=\Theta(\log u)$.

%

\subsection{Random Functions}

\begin{definition}[$k$-wise independent random function]
  A random function $h:[u]\to[r]$ is called \emph{$k$-wise independent} if for any distinct $x_1,\cdots,x_k\in[u]$, and any $y_1,\cdots,y_k\in[r]$,
  \[
	\Pr_h\left[\bigwedge_{i\le k}h(x_i)=y_i\right]=1/r^k.
  \]
\end{definition}

\begin{theorem}
  [\cite{thorup2013simple, christiani2015independence}]
  \label{theorem: random-function}
  Let $[u]$ be a universe, $w=\Omega(\log u)$, $c_1>0$, $r=\poly(u)$, and $k=u^{o(1)}$.
  There exists a data structure for a random function $h:[u]\to[r]$ such that
  \begin{itemize}
	\item with probability $\ge1-1/u$, the data structure is constructed successfully;
	\item upon successful construction of the data structure, $h$ is $k$-wise independent;
	\item the data structure uses space $u^{c_1}$ bits;
	\item for each $x\in[u]$, $h(x)$ is evaluated in $\tilde{O}(1/c_1)$ time in the worst case in the RAM model.
  \end{itemize}
\end{theorem}

\begin{theorem}[Chernoff bound with limited independence \cite{schmidt1995chernoff}]\label{theorem: limited-independence}
  Let $X_1,\cdots,X_n$ be arbitrary $k$-wise independent boolean random variables with $\Pr[X_i=1]=p$ for any $i\in[n]$.
  Let $X\triangleq\sum_iX_i,\mu\triangleq\mathbb{E}[X]=np$, then for any $\delta>0$, it holds that
  \[
	\Pr\left[\,X\ge(1+\delta)\mu\,\right]\le\exp(-\mu\delta^2/2),
  \]
  as long as $k\ge\lceil\frac{\mu\delta}{1-p}\rceil$.
\end{theorem}


%

\subsection{Adaptive Prefixes}

Given a sequence $S=(x_1,x_2,\ldots)$ of strings,
let $\alpha_m(S)=\{\alpha_m(x_1), \alpha_m(x_2), \ldots\}$ be a collection of prefixes, such that for every $x_i\in S$, the $\alpha_m(x_i)$ is the shortest prefix, of length at least $m$, of the binary representation of $x_i$, such that $\alpha_m(x_i)$ prefixes no other $x_j\in S$.
Note that for any string $y$, there is at most one $x\in S$ such that $\alpha_m(x)\sqsupseteq y$ as long as $\alpha_m(S)$ exists.
In particular, $\alpha_m(S)$ does not exist if there are $i\ne j$ such that $x_i= x_j$.

The prefixes are stored in lexicographical order, thus we refer $k$-th prefix as the prefix with rank $k$ in lexicographical order.

\begin{theorem}
  [Refined from \cite{bender2017bloom,bender2018bloom}]
  \label{theorem: adaptive-remainders}
  Let $c_0,c_1>1$ be two constants where $c_0>c_1$.
  For a random sequence $S=(x_1,\cdots)$ of strings drawn from $(\HammingCube{c_0\log u})^{\le c_3\log u}$ uniformly at random with replacement, with probability at least $1-u^{-c_1}$, the prefix collection $\alpha_{\log\log u}(S)$ exists and can be represented with at most $c_2\log u$ bits, where $c_2>0$ is determined by $c_1,c_3$.
  Furthermore, the following operations are supported in constant time:
  \begin{itemize}
	\item $\mathrm{insert}(y)$: update the representation by inserting a new string $y\in\HammingCube{c_0\log u}$ to $S$, when there is at most one $x\in S$ such that $\alpha_{\log\log u}(x)\sqsupseteq y$;
	\item $\mathrm{lookup}(y)$: given any query $y\in\HammingCube{c_0\log u}$, return the rank of the only $z\in\alpha_{\log\log u}(S)$ that prefixes $y$, and return \texttt{NO} if there does not exist such a $z$;
	\item $\mathrm{lowerbound}(y)$: given any query $y\in\HammingCube{\log\log u}$, return the lowest rank of all $z\in\alpha_{\log\log u}(S)$ that $z\sqsubseteq y$, and return $0$ if there is no $z\sqsubseteq y$ in the collection;
  \end{itemize}
\end{theorem}
For completeness, we prove it in Appendix \ref{proof: adaptive-remainders}.

\ifx\mainfile\undefined
\bibliography{refs}
\bibliographystyle{alpha}

\end{document}
\fi

\ifx\mainfile\undefined
\documentclass[11pt]{article}
\input{macro}

\begin{document}
\fi

\section{Data Structures for Sets of Unknown Sizes}
In this section, we present our filter and dictionary data structures for sets of unknown sizes.

\subsection{The Succinct Dynamic Filters}
The following theorem is a formal restatement of Theorem~\ref{main-thm:filter-informal}.

\begin{theorem}
  [Dynamic filter - formal]
  \label{main-theorem-via-prefix-matching}
  Let $0<\epsilon<1$, $[u]$ the data universe, and $\delta= u^{-C}$, where $C>1$ is an arbitrary constant. Assume the word size $w=\Theta(\log u)$.
  There exists a data structure for approximate membership for subsets of unknown sizes of $[u]$, such that
  \begin{enumerate}
	\item 
	for any $n=\omega(\log u)$ and $n<u$, the data structure uses $n(\log(1/\epsilon)+\log\log n+O(\log\log\log u))$ bits of space after insertions of any $n$ key, and extra $u^{c}$ precomputed bits that are independent of the input, where $0<c<1$ is an arbitrary small constant;
	\item each insertion and membership query takes $O(1)$ time in the worst case; 
	\item after each insertion, a failure may be reported by the data structure with some probability, and for any sequence of insertions, the probability that a failure is ever reported is at most~$\delta$, where the probability is taken over the precomputed random bits;
	\item conditioned on no failure, each membership query is answered with false positive rate at most $\epsilon$.
  \end{enumerate}
\end{theorem}


%
%
%
%

As we mentioned in the introduction, our data structure has $\log n$ stages when handling $n$ insertions.
The $i$-th stage is from the insertion of the $(2^{i-1}+1)$-th key to the $2^i$-th key -- the database size doubles after each stage.

The main strategy is to reduce the problem of (approximate) membership to (exact) \emph{prefix matching}.
More formally, in the \emph{prefix matching} problem, we would like to maintain a set of binary strings $\{s_1,s_2,\ldots\}$ of possibly different lengths, supporting
\begin{itemize}
  \item insert($s$): add string $s$ to the set;
  \item query($y$): decide of any string $s$ in the set is a prefix of $y$.
\end{itemize}

To this end, our filter first applies a \emph{global} hash function $h$ such that $h:[u]\to[u^{c_2}]$ is $(c_1\log u)$-wise independent according to Theorem \ref{theorem: random-function}, where $c_1>0$ is a constant to be fixed later, and $c_2$ is a sufficiently large constant (which in fact, is the $c_0$ in Theorem~\ref{theorem: adaptive-remainders}).
To insert a key $x$ in stage $i$, we calculate its hash value $h(x)$, and then insert the $\ell_i$-bit prefix of $h(x)$, for some parameter $\ell_i$.
To answer a membership query $y$, we simply calculate $h(y)$ and search if any prefix of $h(y)$ is in the database.
If no prefix of $h(y)$ is in the database, we output \texttt{NO}; otherwise, we output \texttt{YES}.
It is easy to see that this strategy will never output any false negatives.
On the other hand, by union bound, if the query $y$ is not in the set, the probability that the query algorithm outputs \texttt{YES} is at most
\[
  \sum_{i=1}^{\log u}2^{i}\cdot 2^{-\ell_i},
  \]
since $h$ is $(c_1\log u)$-wise independent (in particular, it is pairwise independent), then the probability that the $\ell_i$-bit prefix of $h(y)$ matches with the prefix of the hash value $h(x)$ of key $x$ is $2^{-\ell_i}$.
Hence, by setting
\begin{align}
\ell_i\triangleq i+\log(1/\epsilon)+\log i+\log\log\log u+2,\label{eq:def-l-i}
\end{align}
the false positive rate is at most
\[
  \sum_{i=1}^{\log u}2^{i}\cdot 2^{-i-\log (1/\epsilon)-\log i-\log\log\log u-2}=\epsilon\cdot \sum_{i=1}^{\log u}\frac{1}{4i\log\log u}<\epsilon.
\]

 We use $\cD_{c_1\log u}$ to denote the distribution of the random insertion sequence $y_1,y_2,\ldots,y_n$ for prefix matching constructed above.
 Formally, $\cD_{c_1\log u}$ is the distribution of a sequence of random strings $y_1,y_2,\ldots,y_n$ obtained from $(c_1\log u)$-wise independent sequence $z_1,z_2,\ldots,z_n\in[u^{c_2}]$ by truncating: $\forall 1\le j\le n$, $y_j=(z_j)_{\le\ell_i}$, where $i={\lceil\log j\rceil}$.

\begin{lemma}
  [Prefix matching]
  \label{DS: unknown-prefix-matching}
  Let $\delta=u^{-C}$, where $C>1$ is an arbitrary constant.
  There exist a constant $c_1$ and a deterministic data structure for prefix matching such that
  \begin{enumerate}
	\item for any $n= \omega(\log u)$ and $n<u$, the data structure uses $n(\ell_{\lceil\log n\rceil}-\log n+O(\log\log\log u))$ bits of space after $n$ insertions, and extra $u^{c}$ precomputed bits, where $0<c<1$ is an arbitrary small constant;
	\item each insertion and query takes $O(1)$ time in the worst case;
	\item after each insertion, a failure may be reported by the data structure, and for a random sequence of insertions drawn from $\cD_{c_1\log u}$, the probability that a failure is ever reported is at most $\delta$, where the probability is taken over $\cD_{c_1\log u}$;
  \item every query is answered correctly if no ``fail'' is reported.
  \end{enumerate}
\end{lemma}

We present the construction in Section~\ref{section: prefix-matching}.
Using this prefix matching data structure, the space usage of the filter is
\begin{itemize}
  \item $n(\ell_{\lceil \log n\rceil}-\log n+O(\log\log\log u))=n(\log (1/\epsilon)+\log\log n+O(\log\log\log u))$ bits,
  \item and $u^c$ bits for storing $h$ by Theorem~\ref{theorem: random-function} and for the precomputed lookup tables described in Appendix~\ref{section: compute RAM}, both independent of the operation sequence.
\end{itemize}
Each insertion and query can be handled in constant time given the data structure does not fail. 
This proves Theorem~\ref{main-theorem-via-prefix-matching}.

\subsection{The Succinct Dynamic Dictionaries}
\label{section: dictionaries}
The data structure for prefix matching also works well as a dictionary data structure for the insertions with keys are sampled uniformly at random.
A worst-case instance can be converted into a random instance by a random permutation 
 $\pi:[u]\to[u]$.
Assuming an idealized $(c_1\log u)$-wise independent random permutation whose representation and evaluation are efficient, the data structure for prefix matching in Lemma~\ref{DS: unknown-prefix-matching} can be immediately turned to a dictionary.
However, the construction of $k$-wise independent random permutation with low space and time costs is a longstanding open problem~\cite{kaplan2009derandomized}. 

We show that our data structure can solve the dictionary problem in the worst case unconditionally, at the expense of extra $u^c$ bits of space for storing random bits which are independent of the input.
\begin{theorem}
  [Dynamic dictionary - formal]
  \label{main-theorem-dictionary}
  Let $[u]\times\HammingCube{v}$ be the data universe, and $\delta= u^{-C}$, where $C>1$ is an arbitrary constant. Assume the word size $w=\Theta(v+\log u)$. 
  There exists a data structure for dictionary for sets of unknown sizes of key-value pairs from $[u]\times\HammingCube{v}$, such that
  \begin{enumerate}
	\item for any $n=\omega(\log u)$ and $n<u$, the data structure uses $n(\log(u/n)+v+O(\log\log\log u))$ bits of space after insertions of any $n$ key-value pairs, and extra $u^{c}$ precomputed bits that are independent of the input, where $0<c<1$ is an arbitrary small constant;
	\item each insertion and query takes $O(1)$ time in the worst case; 
	\item after each insertion, a failure may be reported by the data structure with some probability, and for any sequence of insertions, the probability that a failure is ever reported is at most~$\delta$, where the probability is taken over the precomputed random bits;
	\item conditioned on no failure, each query is answered correctly.
  \end{enumerate}
\end{theorem}

The details of the data structure are postponed to Secion \ref{section: unconditionaly dictionary}. 

\ifx\mainfile\undefined
\bibliography{refs}
\bibliographystyle{alpha}

\end{document}
\fi

\ifx\mainfile\undefined
\documentclass[11pt]{article}
\input{macro}

\begin{document}
\fi

\section{Prefix Matching Upper Bound}
\label{section: prefix-matching}
In this section, we prove Lemma~\ref{DS: unknown-prefix-matching}.

Recall the distribution $\cD_{c_1\log u}$ of random insertion sequence $y_1,y_2,\ldots,y_n$ assumed in Lemma~\ref{DS: unknown-prefix-matching}.
Given an insertion sequence $\bar{y}=(y_1,y_2,\ldots,y_n)\sim\cD_{c_1\log u}$, we define the \emph{core set} $B(\bar{y})\triangleq\{x\in \bar{y}:\forall x'\in \bar{y}, x=x'\lor x'\not\sqsupseteq x\}$, and its subset $B^{(a,b]}\triangleq\{x\in B:|x|\in(a,b]\}$ for any $a<b$.
Let $\cD^{(a,b]}_{c_1\log u}$ denote the distribution of $B^{(a,b]}$.
We say that a random sequence $Y$ of strings is drawn from $\cD^{(a,b]}_{c_1\log u}$ if it can be obtained by permuting the random core set $B^{(a,b]}$. 

We show that Lemma \ref{DS: unknown-prefix-matching} is true as long as there exist a family of  deterministic data structures for prefix matching with known capacity $m$.
An instance of the data structure $D=D(m,\ell)$ is parameterized by capacity $m<u$, and string length upper bound $\ell\ge\log m$.
The data structure uses $u^c$ bits extra space whose contents  are precomputed lookup tables, and supports following functionalities with good guarantees:
\begin{itemize}
  \item $\PMInit{D}$ and $\PMDestr{D}$: 
	subroutines for initializing and destroying $D$ respectively.
	The data structure is successfully initialized (or destroyed) after invoking $\PMInit{D}$ ($\PMDestr{D}$) consecutively for $O(m)$ times. 
	When successfully initialized, $D$ uses space $O(m)$ bits. The
	$\PMInit{D}$'s are invoked before all other subroutines and  $\PMDestr{D}$'s are invoked after all other subroutines.
%
  \item $\PMInsert{D}{x}$:  insert string $x$ to $D$, where $\log m<|x|\le\ell$.
	After $n$ insertions, $D$ uses at most $n(\ell-\log m+2\log\log\log u)+O(m)$ bits.
	Each insertion may cause $D$ to fail.
  A failure ever occurs for a random insertion sequence $Y$ with probability at most $u^{-2C}$, as long as $Y$ is drawn from $\cD^{(\log m,\ell]}_{c_1\log u}$, where $c_1$ is suitably determined by constant $C$.
\item $\PMQuery{D}{x}$:  return one bit to indicate whether there exists a prefix of $x$ in $D$. The correct answer is always returned as long as $D$ has not failed.
\item $\PMDecr{D}$:  try to delete an arbitrary string $y$ in $D$ and return the $y$ if $y$ is deleted.
An invoking may delete nothing and hence nothing is returned, but it guarantees that the total number of such empty invoking is at most $m$.
  Each invoking that successfully deletes a string frees space $\ell-\log m$ bits.
  The $\PMDecr{D}$'s are invoked after all insertions.
\end{itemize}

\begin{claim}
  \label{claim: interfaces}
  Given the deterministic data structures supporting above functionalities in  constant time in the worst case, Lemma \ref{DS: unknown-prefix-matching} is true.
\end{claim}

\begin{proof}
  We use an auxiliary structure called \emph{truth table} to deal with short strings.
  A truth table $T_i$ is a bitmap (i.e. array of bits) of length $2^i$ and supports the required functionalities in the worst cases:
  \begin{itemize}
	\item $T_i$ is initialized to the all-0 string $0^{2^i}$,
	where each invoking of $\PMInit{T_i}$ extends $T_i$ by one 0 until $T_i$ is of length $2^i$, and each invoking of $\PMDestr{T_i}$ shrinks $T_i$ by one bit until $T_i$ is fully destroyed;
	\item to insert $x$ where $|x|=i$, we set $T_i[x+1]\gets 1$;\footnote{For $A$, a list or array of items, we let $A[i]$ denote the $i$-th item of $A$.}
	\item to query $x$ where $|x|=i$, we return \texttt{YES} if $T_i[x+1]=1$ and return \texttt{NO} if otherwise;
	\item to decrement $T_i$, we maintain a $j$ that traverses from $1$ to $2^i$, and at each time set $T_i[j]\gets 0$, return $j-1$ if $T_i[j]=1$, and increment $j$ by 1.
  \end{itemize}
  

  Initially, the prefix matching data structure required by Lemma~\ref{DS: unknown-prefix-matching} consists of $T_0,T_1$ and $D_0=D(1,\ell_0),D_1=D(2,\ell_1)$ respectively with capacities $1,2$, and string lengths $\ell_0,\ell_1$, where $\ell_i$ is defined in~Eq\eqref{eq:def-l-i}.

  To insert $x$, which is the $n$-th insertion, we set $i\gets \lceil\log n\rceil$, invoke $\PMInsert{D_i}{x}$ if there is no prefix of $x$ has been inserted. 
  Then we execute the following procedure for $10$ times to maintain our data structure:
  \begin{enumerate}
	\item 
	  If $T_{i-1}$ is non-empty, we decrement it by invoking $\PMDecr{T_{i-1}}$.
	  If a $y$ is returned, we insert it into $T_i$ by invoking $\PMInsert{T_i}{y\circ 0}$ and $\PMInsert{T_i}{y\circ 1}$.
	\item 
	  If $D_{i-1}$ is non-empty, we invoke $\PMDecr{D_{i-1}}$.
	  If a $y$ is returned, we insert it into $D_i$ by invoking $\PMInsert{D_i}{y}$ when $|y|>i$ and insert $y$ into $T_i$ by invoking $\PMInsert{T_i}{y}$ otherwise.
	\item
	  If $T_{i-1}$ (or $D_{i-1}$) is empty but not destroyed yet, we invoke $\PMDestr{T_{i-1}}$ (or $\PMDestr{D_{i-1}}$).
	\item
	  If $T_{i-1}$ (or $D_{i-1}$) has been destroyed, we invoke $\PMInit{T_{i+1}}$ (or $\PMInit{D_{i+1}}$ for $D_{i+1}=D(2^{i+1},\ell_{i+1})$ with capacity $2^{i+1}$ and string length upper bound $\ell_{i+1}$), where $\ell_i$ is defined in~Eq\eqref{eq:def-l-i}.
  \end{enumerate}
  A failure is reported whenever a failure is reported during insertion to $D_i$.
  
   Clearly, for any integer $n\in [2^i,2^{i+1})$, after $n$ insertions, all inserted strings are stored in either $D_{i-1},T_{i-1}$ or $D_i,T_i$.
	 By the time $n$ reaches $2^{i+1}$, $D_{i+1},T_{i+1}$ have been initialized, all inserted strings are stored in $D_i,T_i$, and $D_{i-1},T_{i-1}$ have been destroyed. 
	
  Consider the insertion sequence for a fixed $D_i$.
  Observe that the strings inserted into $D_i$ must be in the core set $B^{(i,\ell_i]}(\bar{y})$. 
  Therefore the insertion sequence is drawn from $\cD^{(i,\ell_i]}_{c_1\log u}$, which means that insertions to each $D_i$ ever failed with probability at most $\delta$.  
	By union bound, a failure is ever reported with probability at most $\sum_{i=1}^{\log u} u^{-2C}\le u^{-C}=\delta$.

	Overall, the data structure uses at most $n(\ell_{\lceil\log n\rceil}-\log n+O(\log\log\log u))\le n(\log(1/\epsilon)+\log\log n+O(\log\log\log u))$ bits after $n$ insertions, besides the $u^c$ precomputed bits.

	Suppose $n$ strings has been inserted, let $i\gets \lceil\log n\rceil$.
	To query $x$, we invoke $\PMQuery{D_{i-1}}{x}$, $\PMQuery{D_{i}}{x}$,  $\PMQuery{T_{i-1}}{x_{\le i-1}}$, $\PMQuery{T_{i}}{x_{\le i}}$ simultaneously, and return \texttt{YES} if any one of the invokings returns \texttt{YES}.

	Obviously each insertion and query takes constant time in the worst case, and it is easy to check that every query is correctly answered as long as no failure is reported.
\end{proof}


\section{Succinct Prefix Matching with Known Capacity}
\label{section: known-prefix-matching}
We now describe the data structures required by Claim~\ref{claim: interfaces}.
The pseudocodes are given in Appendix \ref{section: implementation}.


The data structure consists of a main table and $m/\log u$ subtables.

We partition each binary string $x$ into four consecutive parts:
$st(x), hd(x), hs(x), rt(x)$ of lengths $\log(m/\log u),\log(\log u/\log \log u),\log\log \log u, |x|-\log m$ respectively.
Roughly speaking, a datapoint $x$ will be distributed into a subtable according to $st(x)$, then be put into a data block of size $\log u/\log \log u$ according to the order it is inserted, therefore we can save $|st(x)|+|hd(x)|-O(1)$ bits for each datapoint by properly encoding.

\begin{table}[ht]
  \begin{center}
	\begin{tabular}{|c|l|c|}
	  \hline
	  \multirow{1}{*}{input} 
	  & input independence & $c_1\log u$\\
	  \hline
	  \hline
	  \multirow{2}{*}{main table} 
	  & main table length & $m/\log u$\\
	  & max-load & $c_3\log u$\\
	  \hline
	  \hline
	  \multirow{3}{*}{subtable} 
	  & max-load on $hd(x)\circ hs(x)$ & $c_4\log u/\log \log u$\\
	  & capacity of data block & $\log u/\log \log u$\\
	  & failure probability of fingerprint collection & $u^{-c_5}$\\
	  \hline
	\end{tabular}
  \end{center}
  \caption{The setting of parameters for the data structure.}
\end{table}

\subparagraph{Main Table.}
The main table consists of $m/\log u$ entries, each of which contains a pointer to a subtable.
Each insertion/query $x$ is distributed into an entry of the main table addressed by $st(x)$.
Recall the word size $w=\Theta(\log u)$.
The main table uses $mw/\log u=O(m)$ bits.

Recall that $\bar{y}=(y_1,y_2,\ldots,y_n)\sim\cD_{c_1\log u}$ is transformed from a $(c_1\log u)$-wise independent sequence $Z=(z_1,z_2,\cdots,z_n)$ by truncating.
The insertion sequence $Y$ is drawn from $\cD^{(\log m,\ell]}_{c_1\log u}$ by permuting $B^{(\log m,\ell]}$, the restriction of the core set $B(\bar{y})$ to the strings whose lengths ranges within $(\log m,\ell]$.

Let $Y_i,Z_i$ denote the subsequences of $Y,Z$ which contain all the strings that have prefix~$i$, respectively.
By definitions, $|Y_i|\le |Z_i|$.
%
%
%
Recall that $Z$ are $(c_1\log u)$-wise independent.
Due to Theorem \ref{theorem: limited-independence}, the load of entry $i$ exceeds $c_3\log u$ with probability
\begin{align}
  \Pr\left[\,|Y_i|\ge c_3\log u\,\right]\le\Pr\left[\,|Z_{i}|\ge c_3\log u\,\right]\le\exp(-(c_3-1)^2\log u/2),
\end{align}
as long as $c_1\ge\lceil2(c_3-1)^2\rceil$.
Therefore the max-load of entries of the main table is upper bounded by $c_3\log u$ with probability at least $1-(m/\log u)\exp(-(c_3-1)^2\log u/2)$.
The data structure reports failure if any entry of the main table overflows.
In the rest of the proof, we fairly assume $|Y_i|\le |Z_{i}|\le c_3\log u$ for all $i$.

Observe that a datapoint $x$ can be identified with $hd(x)\circ hs(x)\circ rt(x)$ if the entry $i=st(x)$ it is distributed into is fixed.
Therefore we let $Y'_i,Z'_i$ denote the subsequences generated from $Y_i,Z_i$ by discarding the left-most $\log(m/\log u)$ bits.


\subparagraph{Subtable.}
Each subtable $i$ consists of the following parts to be specified later:
\begin{itemize}
  \item a collection of fingerprints $\alpha_{\log\log u}(Y'_i)$ and its indicator list $I_i$;
  \item an (extendable) array of navigators $N_i$;
  \item an (extendable) array of data blocks $A_i$;
  \item two buffers, $B_{i,u},B_{i,r}$;
  \item constant many other local variables.
\end{itemize}
All the datapoints are stored in array $A_i$.
Given a datapoint $x$, it is easy to see that the addresses of the entries which contains the information of $x$ is high correlated with the order it is inserted, since any insertion takes constant time in the worst case.
Hence we take the fingerprints $\alpha_{\log\log u}(Y'_i)$, indicators $I_i$, navigators $N_i$, and a tricky way to encode a data block as clues to locate the entries which maintain $x$.  
Recall that new insertions is put into the latest data block using a dynamic construction, and we reorganize the full dynamic data block into a static construction.
We use buffer $B_{i,u}$ to maintain the dynamic block, and use buffer $B_{i,r}$ to ``de-amortize'' the reorganization.

At first consider a static version of our data structure.
In the static version, the buffers and the indicator list are unnecessary.
Let $n_i\triangleq|Y_i|$. 

\subparagraph{Fingerprints.}
The collection of fingerprints $\alpha_{\log\log u}(Y'_i)$ is obtained by applying Theorem \ref{theorem: adaptive-remainders} on $Y'_i$ with guarantee $c_1\ge c_3$. 
Note that $Z'_{i}$ are mutually independent as long as $c_1\ge c_3$.
Due to Theorem \ref{theorem: adaptive-remainders}, there exists a constant $c''>0$ such that a fingerprint collection $\alpha_{\log\log u}(Z'_i)$ for $Z'_{i}$ can be represented in $c''\log u$ bits with probability $1-u^{-c_5}$.

We show that there exists a injective function $P:[|Y'_i|]\to[|Z'_i|]$ such that $\forall j\in[|Y'_i|], Y'_i[j]\sqsupseteq Z'_i[P(j)]$.
Due to the injective function $P$ and the guarantee that $Y$ is prefix-free, the fingerprint collection of $Y'_i$ can be represented with the same space and probability guarantees as above.

We define $P:[|Y_i|]\to[|Z_i|]$ by  $P(j)\triangleq\min\{k\in[|Z_i|]:Y_i[j]\sqsupseteq Z_i[k]\}$.
By the definition, for any $y\in Y_i$, there is $z\in Z$ such that $y\sqsupseteq z$.
Recall that all the strings in $Y_i$ has prefix $i$.
Hence for any $y\in Y_i, z\in Z$ such that $y\sqsupseteq z$, it holds that $i\sqsupseteq z$, i.e. $z\in Z_i$.
Thus for any $j\in[|Y_i|],\{k\in[n]:Y_i[j]\sqsupseteq Z_i[k]\}\ne\emptyset$.
Therefore $P$ is well-defined.
On the other hand, for distinct $j,l\in[|Y_i|]$, $\{k\in[n]:Y_i[j]\sqsupseteq Z_i[k]\}$ and $\{k\in[n]:Y_i[l]\sqsupseteq Z_i[k]\}$ are disjoint, since $Y_i$ is prefix-free and there is no such $z$ that $x,y$ prefix $z$ simultaneously for distinct $x,y\in Y_i$.
Therefore $P$ is injective.
Recall that $Y'_i,Z'_i$ are generated by removing the prefix $i$ from the strings in $Y_i,Z_i$: $\forall j, Y_i[j]=i\circ Y'_i[j], Z_i[j]=i\circ Z'_i[j]$.
Therefore $P$ works for $Y'_i,Z'_i$ too, i.e.~$\forall j\in[|Y'_i|], Y'_i[j]\sqsupseteq Z'_i[P(j)]$.

A failure is reported if any fingerprint collection can not be represented within $c''\log u$ bits, which occurs with probability at most $u^{-c_5}$.

The fingerprints are sorted  lexicographically, so that by the $j$-th fingerprint  we mean the $j$-th in lexicographical order. 
For simplicity, we write $\alpha_i\triangleq\alpha_{\log\log u}(Y'_i)$.



A failure is reported if there are more than $c_4\log u/\log\log u$ datapoints share identical $hd(x)$ and $hs(x)$, which occurs with probability at most
\begin{align}
  \binom{c_3\log u}{c_4\log u/\log \log u}(\frac{1}{\log u})^{c_4\log u/\log \log u}<u^{-(c_4-0.01)}.
  \label{eq: hd-hs load}
\end{align}
The fingerprints cost $O(\log u)$ bits per subtable if no failures.

\subparagraph{Navigators.}
$N_i$ is an array of pointers.
For any datapoint, the rank of its fingerprint is synchronized with the index of its navigator. 
In particular, for the $k$-th fingerprint in $\alpha_i$, $N_i[k]$ is the address of the data block which maintains the datapoint with the fingerprint.
A data block maintains up to $\log u/\log \log u$ datapoints, thus there are at most $c_3\log \log u$ data blocks.
The navigators cost at most $n_i\log\log \log u+O(n_i)$ bits of space.


\subparagraph{Data Blocks.}
$A_i$ is interpreted as an array of data blocks, with each data block holding up to $\log u/\log \log u$ datapoints.

Consider the following succinct binary representation (called \emph{pocket dictionary} in \cite{bercea2019fully}) 
of a collection of datapoints $F\in\binom{\HammingCube{\ell}}{m}$:
The representation consists of two parts $header(F)$ and $body(F)$.
Let $header(x), body(x)$ denote the left-most $\log m$ bits and the right-most $\ell-\log m$ bits of $x$.
Let $n'_i\triangleq|\{x\in F|header(x)=i\}|$, and $F=(x_1,\cdots,x_m)$ sorted lexicographically.
Then $header(F)\triangleq 0\circ1^{n'_0}\circ 0\circ 1^{n'_1}\circ 0\cdots1^{n'_{m-1}}$ and $body(F)\triangleq body(x_1)\circ body(x_2)\cdots\circ body(x_m)$.
It is easy to see that this representation uses $2m+m(\ell-\log m)$ bits of space.

Our static data block is a variant of this representation.
Let $(x_1,\cdots,x_{\log u/\log \log u})$ be the sorted list of datapoints maintained by  data block $j$.
Data block $j$ consists of a list of headers $(hd(x_1),\cdots,hd(x_{\log u/\log \log u}))$, a list of identities $(hs(x_1),\cdots,hs(x_{\log u/\log \log u}))$ and an array of the rest part of datapoints $(rt(x_1),\cdots,rt(x_{\log u/\log \log u}))$.
The header list are represented in the same way as in the pocket dictionary, the identity list is the concatenation $hs(x_1)\circ hs(x_2)\cdots\circ hs(x_{x/\log \log u})$, and the rest part array is the concatenation $\PFC(rt(x_1))\circ \PFC(rt(x_2))\cdots\circ \PFC(rt(x_{x/\log \log u}))$, where $\PFC(\cdot)$ is the prefix-free code in Claim \ref{claim: prefix-free-code}.

Recall that $|hd(x)|=\log(\log u/\log \log u),|hs(x)|=\log\log \log u$.
Therefore the data blocks for the subtable cost at most $O(n_i)+n_i(\ell-\log m+\log\log \log u)$ bits of space.

\subparagraph{Query in a Static Data Block.}
Recall that the fingerprints are sorted in lexicographical order, and the indices of the navigators are synchronized with the ranks of corresponding  fingerprints. 
Also note that a navigator costs $\log\log\log u+O(1)$ bits, there are at most $c_4\log u/\log\log u$ datapoints share $hd(x)\circ hs(x)$ with any query $x$.
By putting everything together, we can retrieve all the navigators of the datapoints which have the same $hd(x)\circ hs(x)$ with query $x$, and learn a $k'$ such that the unique suspected datapoint is the $k'$-th datapoint among the datapoints which share the $hd(x)\circ hs(x)$ in its data block.
On the other hand, we can retrieve the header list and identity list with constant number of memory accesses.
Consequently, we can retrieve the rest part of the suspected datapoint efficiently.
See the pseudocode in Appendix \ref{section: implementation of query} for more details.


The space usage is upper bounded by $m+(m/\log u)\cdot O(\log u)+n\log \log \log u+O(n)+n(\ell-\log m+\log\log \log u)\le n(\ell-\log m+2\log\log \log u)+O(m)$ bits.
And the query time is clearly constant.

\subparagraph{Insertion.}
The new insertions will be put into a data block under construction temporarily, and the data block will be reorganized to the static version as long as the data block is full (which means, there are $\log u/\log \log u$ datapoints stored in it).
A data block under construction consists of two incomplete lists for headers and identities, and an (extendable) array of the rest parts of datapoints.
Note that the space usages of incomplete lists are identical with the ones of the complete lists, it wastes at most $O(\log u)$ bits per dynamic data block.

Reorganizing a data block (i.e.~sorting a data block) can be expensive, therefore we finish this work during the procedure that a new data block under construction is being filled.
Hence there are two dynamic data blocks, one under construction and one under reorganization, besides the static ones.

Note that the collection of fingerprints $\alpha_i$ can be updated  dynamically with small costs.
To retrieve the datapoint $y$ with fingerprint $\alpha(y)$, the only things we need are the address of the data block which maintains $y$ and the in-block index of $y$.
(recall that $hd(y)$ and $hs(y)$ are known due to the fingerprint collection.)
It is easy to learn the address as long as we know that $y$ is in a dynamic data block, since there are at most two dynamic blocks.
We use the buffers to record the in-block index, and use the indicators to inform whether $y$ is in a dynamic data block.

The list of indicators is a string from $\{1,2,3\}^{n_i}$.
The value of $i$-th indicator implies which kind of data block the datapoint corresponding to $i$-th fingerprint is stored in.
For a static data block, the query algorithm works in previous way.
For a dynamic data block, the address of the data block can be easily learnt with the counter $n_i$.

The two buffers are arrays of pointers from $[\log u/\log \log u]^{\log u/\log \log u}$, so they fit in constant number of  memory cells.
In particular, for a indicator $I_i[j]$ which is the $k$-th indicator has value $2$ (or $3$), $B_{i,u}[k]$ (or $B_{i,r}[k]$) is the in-block index of the rest part which corresponds to $j$-th fingerprint.

The new insertion is not a prefix of some preceded insertion due to our guarantees. 
Therefore, to insert $x$, we simply append $hs(x),rt(x)$ to the identity list and rest part array, update the header list, fingerprint collection, and indicator list, and insert a proper pointer into $B_{i,u}$, which overall takes constant time. 
And to query $x$ in a dynamic data block, where $x$ corresponds to a datapoint $y$ stored in the data block, we need to retrieve the address of the data block with counter $n_i$ and the in-block index of $y$ with the buffers, then retrieve $rt(y)$.
See the pseudocodes in Appendices~\ref{section: implementation of insertion} and~\ref{section: implementation of query}  for more details.


\subparagraph{Reorganizing a Data Block.}
%
The reorganization procedure starts as long as the data block under construction is full.
Informally, the reorganization procedure works as follows:
\begin{enumerate}
  \item Update the list of indicators and copy $B_{i,r}\gets B_{i,u}$. (We guarantee that the preceded reorganization process has been finished before the buffer $B_{i,u}$ is full)
  \item Insert the address of the data block in proper positions of the navigator list.
  \item Sort the array of rest parts and the list of identities according to the pointers in the buffer $B_{i,r}$ while keep the buffer updated.
	Note that the sorting can be done within time cost $O(\log u/\log \log u)$: we enumerate $j\in[\log u/\log\log u]$, find $j'$ such that $B_{i,r}[j']=j$, swap $hs_j,rt_j,B_{i,r}[j]$ with $hs_{B_{i,r}[j]},rt_{B_{i,r}[j]},B_{i,r}[j']$ one by one, where $hs_j,rt_j$ is the $j$-th item of the corresponding list and array.
	\item  Update indicator list.
\end{enumerate}
The total time cost is $O(\log u/\log \log u)$, which can be simulated by $\log u/\log \log u$ operations, each costing $O(1)$ time, within a data block.
See Appendix \ref{section: implementation of insertion} for more details.

The two dynamic data blocks waste at most $O(\log u)$ bits on the two incomplete lists for headers and identities, therefore we uses at most extra $O(m)$ bits of space.

%
%

\subparagraph{Setting the Constant Parameters.}
Our data structure may fail at the load balancing on subtables, constructing the fingerprint collections for subtables, and the load balancing on headers of fingerprint collections.
By the union bound, the failure probability is at most
\begin{align}
  \frac{m}{\log u}\big(\exp(-(c_3-1)^2\log u/2)+u^{-c_5}\big)+m2^{-(c_4-0.01)\log u},
  \label{eq: failure probability}
\end{align}
when $c_1\ge\max\{\lceil2(c_3-1)^2\rceil,c_3\}$.
Since $m<u$ ,the failure probability can be as small as $\delta=u^{-2C}$ if we set the constants $c_1,c_3,c_4,c_5$ to be sufficiently large.

\bigskip
The initialize, decrement, and destroy subroutines are easy to implement, which are postponed to Appendix \ref{section: implementation}.

\ifx\mainfile\undefined
\bibliography{refs}
\bibliographystyle{alpha}
\end{document}
\fi

\ifx\mainfile\undefined
\documentclass[11pt]{article}
\input{macro}

\begin{document}
\fi

\section{Unconditional Succinct Dictionary}
\label{section: unconditionaly dictionary}

We show that our data structure can solve the dictionary in the worst case unconditionally.
In our data structure for prefix matching, the randomness is used only for:
\begin{enumerate}
  \item load balancing on the subtables;
  \item the representation of the adaptive prefixes;
  \item load balancing on the $hd(x)\circ hs(x)$'s.
\end{enumerate}
Note that we should decode $st(x)$ from subtable index $i$, decode $hd(x)\circ hs(x)$ from bucket index in adaptive prefixes, and decode $hd(x)$ from bucket index in data block.
To achieve the identical guarantees with prefix matching, we apply a weaker but strong enough random permutation.

\begin{definition}
  [Feistel permutation]
  Given any $x\in[u]$, let $x_L, x_R$ respectively denote the $\log m$ left-most bits and the $\log(u/m)$ right-most bits of the binary representation of $x$, so that $x=x_L\circ x_R$.
  Given any $f:\HammingCube{\log(u/m)}\to\HammingCube{\log m}$, the \emph{Feistel permutation} $\pi_f:[u]\to[u]$ is defined as
  \[
  \pi_f(x)=(x_L\oplus f(x_R))\circ x_R.
  \]
\end{definition}
\noindent
It is easy to verify that $\pi_f$ is indeed a permutation. In fact, $\pi_f(\pi_f(x))=x$ for any $x$ and $f$.

Our dictionary data structure works with three $(c_1\log u)$-wise independent hash functions $f:[\frac{u\log u}{m}]\to[m/\log u]$, $g:[u/m]\to[\log u]$, and $h:[u/m]\to[u^{c_2}]$.
Given a key $x\in[u]$, we let $st'(x)=\pi_f(x), hd'(x)\circ hs'(x)=\pi_g(hd(x)\circ hs(x)\circ rt(x))$ and $hss(x)=h(rt(x))$.
Then we distribute $x$ into subtable $st'(x)$, insert/query $hd'(x)\circ hs'(x)\circ hss(x)$ to the fingerprint collection, and encode $hd'(x)\circ hs'(x)\circ rt(x)$, instead of $hd(x)\circ hs(x)\circ rt(x)$, in its data block.

Consider two datapoints $x,x'$.
If $hd(x)\circ hs(x)\circ rt(x)\ne hd(x')\circ hs(x')\circ rt(x')$, then $st'(x)$ and $st'(x')$ are independent; otherwise $st'(x)\ne st'(x')$.
Therefore for any $i,c$, 
\begin{align}
  \Pr\left[\,|\{x\in Y: st'(x)=i\}|\ge c\,\right]\le\Pr\left[\,|\{x\in Y': st(x)=i\}|\ge c\,\right],
  \label{eq: load upper bound of permutation}
\end{align}
where $Y$ are the insertion sequence, $Y'$ are $(c_1\log u)$-wise independent random insertion sequence.
Hence the load balancing is not worse than the one in the prefix matching case.

Due to Theorem \ref{theorem: adaptive-remainders} and Eq(\ref{eq: load upper bound of permutation}), the fingerprint collection works with the same guarantees.
Clearly $hd(x)\circ hs(x)=(hd'(x)\circ hs'(x))\oplus g(rt(x)), st(x)=st'(x)\oplus f(hd(x)\circ hs(x)\circ rt(x))$, thus the keys can be retrieved precisely.
For the values, we store $rt(x)$ and its value together as a tuple in the data block.

\section{Upper Bounds in Allocate-Free Model}
\label{section: allocate-free}

We mimic the extendable arrays in the allocate-free model.
For simplicity, we modify the navigator list from extendable array to an array of length $c_3\log u$.

Suppose we are dealing with $D_i$.
The main table can be implemented easily since it has fixed length.
Let $s=c_3(\log u)(\log(1/\epsilon)+\log i+O(\log\log\log u))$ be the space usage upper bound of any single subtable.
For a subtable $i$, we maintain a pointers array of length $\lceil\sqrt{s/w}\rceil$ to mimic the extendable array.
Every pointer in the array points to a memory block of $\lceil\sqrt{sw}\rceil$ bits.
Therefore we waste at most 
\[
  (2^i/\log u)\cdot O(w\cdot\sqrt{s/w}+\sqrt{sw})=O(2^i\sqrt{\log(1/\epsilon)+\log i+\log\log\log u})
\]
bits of space.
In conclusion, after $n$ insertions our data structure for filters uses at most
\[
  n(\log(1/\epsilon)+\log\log n+O(\log\log\log u))+O(n\sqrt{\log(1/\epsilon)+\log\log n+\log\log\log u})
\]
bits of space in the allocate-free model.

Similarly, our data structure for dictionaries uses at most
\[
  n(\log (u/n)+v+O(\log\log\log u))+O(n\sqrt{\log (u/n)+v+\log\log\log u})
\]
bits of space after $n$ insertions in the allocate-free model.

\ifx\mainfile\undefined
\bibliography{refs}
\bibliographystyle{alpha}

\end{document}
\fi

\bibliography{refs}

\appendix

\ifx\mainfile\undefined
\documentclass[11pt]{article}
\input{macro}

\begin{document}
\fi

\section{Computing in RAM Model}
\label{section: compute RAM}
\subsection{Basic Subroutines}
Our data structure involves the following computational operations:
\begin{itemize}
  \item $\mathtt{Popcount}(a)$: Given a string of bits $a\in\HammingCube{*}$, return the Hamming weight of $a$ (i.e. return the number of ones in $a$).
  \item $\mathtt{Select}(a,k)$: Given a string of bits $a\in\HammingCube{*}$, and a number $k$, return $i$ such that $a_i$ is the $k$-th one, return $\perp$ if there does not exist at least $k$ ones.
  \item $\mathtt{Extend}(x,y)$: Given a pair of bit strings $x,y\in\HammingCube{*}$ return the shortest prefix $x'$ of $x$ such that $x'\not\sqsupseteq y$; return $x$ if $x\sqsupseteq y$.
  \item $\mathtt{Double}(x)$: Given a bit string $x=\HammingCube{*}$, return $0\circ x_1\circ 0\circ x_2\circ 0\cdots$.
  \item $\mathtt{rDouble}(x)$: Given a bit string $x=\HammingCube{*}$, return $x_2\circ x_4\circ x_6\circ x_8\cdots$.
  \item $\mathtt{Pred}(a,x)$: Given a concatenation of bit string list $a=\perp\circ a_1\circ\perp\circ a_2\circ\perp\circ a_3\circ\perp\cdots$ where $a_1,a_2,\cdots$ are sorted in lexicographical order, and a query $x\in\HammingCube{*}$, return the largest $i$ such that $a_i$ precedes or is equal to $x$ in lexicographical order; return $0$ if all the strings succeeds $x$.
\end{itemize}

It is easy to see that, given $a\in\HammingCube{W}$, $\mathtt{Popcount}(a)$ can be computed in $W/(\log u/c)$ times with a precomputed lookup table using $u^{1/c}\lceil\log(\log u/c)\rceil$ bits of space by partitioning $a$ into $Wc/\log u$ chunks, for any $c>1$.
The lookup table encodes the $\mathtt{Popcount}$ function from $\HammingCube{\log u/c}$ to $[\log u/c]$.

$\mathtt{Extend},\mathtt{Double},\mathtt{rDouble}$ are implemented similarly with lookup tables.

For the $\mathtt{Select}$ subroutine, it can be reduced to a tiny version for $a'\in\HammingCube{\log u/c}$ with the $\mathtt{Popcount}$ subroutine.
Given an $a\in\HammingCube{W}$, we partition it into $W/(\log u/c)$ consecutive substrings of length $\log u/c$, then apply $\mathtt{Popcount}$ on each of them to locate the only substring $j$, which is our $a'$,  which contains the $k$-th one.
Let $k'$ be the sum of the Hamming weight of the substrings precede substring $j$, we apply $\mathtt{Select}$ on $(a',k-k')$.
Again, with a precomputed lookup table using $\tilde{O}(u^{1/c})$ bits of space, the tiny version can be computed in constant number.

For the predecessor search, we also reduce it to tiny version for a concatenation of strings of length at most $\log u/c$.
Given a concatenation of length $W$, we partition it into $W/(\log u/c)$ chunks, then there are at most $W/(\log u/c)$ strings straddle a chunk boundary. 
We retrieve such strings with the $\mathrm{select},\mathrm{rselect}$ operations in the next subsection and the bit shift operations, retrieve their indices with the \textrm{popcount} operation in the next subsection, then match these strings with query $x$ one by one.
For the strings which are entirely contained within a chunk, we let $x'$ be the left-most $\log u/c$ bits of $x$, then run $\mathtt{Pred}$ on the chunks and $x'$ with a precomputed lookup table using $\tilde{O}(u^{2/c})$ bits of space.

In conclusion, all of these subroutines on a string $a\in\HammingCube{W}$ can be implemented in RAM model in $O(Wc/\log u)$ time with precomputed lookup tables using space of $\tilde{O}(u^{2/c})$ bits, for any $c>1$.

\subsection{Non-Trivial Operations} 
The following  non-trivial operations are involved in our algorithms:
\begin{itemize}
  \item $\mathrm{popcount}(a,k)$: return the number of $k$ in array $a$.
  \item $\mathrm{select}(a,k,l)$: return the index of $l$-th $k$ in array $a$.
  \item $\mathrm{rselect}(a,k,l)$: return the index of $l$-th $k$ from the end in array $a$.
  \item $\mathrm{set}(a,k,l)$: set all the $k$ to $l$ in array $a$.
\end{itemize}

Assume $a=(a_1,a_2,\cdots,a_{m'})$ maintains elements from $\HammingCube{\ell}$ and $a$ is represented as $z=0\circ a_1\circ 0\circ a_2\cdots\circ a_{m'}$ without loss of generality.
The assumption wastes at most $O(m)$ bits since the total length of all the arrays in our data structure is $O(m)$.

Let $\mathbf{0}\triangleq 0^{\ell}$, and
\[
  z'_k\triangleq (1\circ \mathbf{0}\circ 1\circ\mathbf{0}\dots)
  \And
  ((z\oplus(0\circ\neg k\circ 0\circ\neg k\circ 0\dots))+(0\circ (\mathbf{0}+1)\circ 0\circ(\mathbf{0}+1)\dots)).
\]

\begin{itemize}
  \item $\mathrm{popcount}(a,k)=\mathtt{Popcount}(z'_k))$.
  \item $\mathrm{select}(a,k,l)=(\mathtt{Select}(z'_k,l)+\ell)/(1+\ell)$.
  \item $\mathrm{rselect}(a,k,l)=(\mathtt{Select}(z'_k,\mathtt{Popcount}(z'_k)+1-l)+\ell)/(1+\ell)$.
  \item $\mathrm{set}(a,k,l)$: $z\gets z'_k\oplus((z'_k\gg\ell)\times (k\oplus l))$.
\end{itemize}

\section{Proof of Theorem \ref{theorem: adaptive-remainders}}\label{proof: adaptive-remainders}

Consider that $x_1,\ldots, x_{c_3\log u}$ are distributed into $2^{\lceil\log \log u\rceil}$ buckets according to their $\lceil\log \log u\rceil$ most significant bits. 
Let $B_j$ denote the $j$-th bucket. 
All $x_i\in B_j$ must have the same prefix $j$.
For each $x_i\in B_j$, define the local prefix $\alpha_j(x_i)$ to be the string obtained by removing the left-most $\lceil\log\log u\rceil$ bits from the global prefix $\alpha(x_i)$, i.e.~for all $x_i\in B_j$, $\alpha(x_i)=j\circ\alpha_j(x_i)$.
$j, \alpha_j(x_i)$ are called the \emph{header}  and \emph{body} of the prefix.
Note that $\alpha_j(x_i)$ may be empty.

To analyze the space usage, consider a random process that we insert the datatpoints one by one as following:
Without loss of generality, suppose we are trying to insert $x$ to bucket $i$, and there are $j$ datapoints in the database. 
At first we find a $y$ such that $\alpha_i(y)$ is a prefix of $x$.
Note that these exists at most one $y$ such that $\alpha_i(y)$ is a prefix of $x$, since the current prefix collection is prefix-free.
If there exists such a $y$, we extend $\alpha_i(y)$ bit by bit until $\alpha_i(y)$ is not a prefix of $x$; otherwise we do nothing.
We then let $\alpha_i(x)$ be the shortest prefix of $x$ such that $\alpha_i(x)$ is not a prefix of all other prefixes, and insert $i\circ\alpha_i(x)$ into the prefix collection.

Let $E(x),A(x)$ denote the number of bits introduced by extending $\alpha_i(y)$ and selecting $\alpha_i(x)$, respectively.
It is easy to see that $E(x)$ has a geometric distribution with parameter $p=1/2$.
Then the sum of $E(x)$'s has a negative binomial distribution, 
\[
  \Pr\left[\,\sum_{x\in X}E(x)\ge cc_3\log u\,\right]\le\binom{(1+c)c_3\log u}{c_3\log u}2^{-cc_3\log u}\le u^{-cc_3+c_3\log(e(1+c))},
\]
for any constant $1<cc_3\le c_0$.
It is not hard to see that $\Pr[A(x)\ge l]\le j/2^l$ for any $l\le c_0$ by the union bound.
Let $l(x)\triangleq A(x)-\lceil\log_2j\rceil$, then $l(x)$ is upper bounded by a random variable which is sampled from a geometric distribution with parameter $1/2$.
Being similar with $E(x)$ case, 
\[
  \Pr\left[\,\sum_{x\in X}l(x)\ge cc_3\log u\,\right]\le\binom{(1+c)c_3\log u}{c_3\log u}2^{-cc_3\log u}\le u^{-cc_3+c_3\log(e(1+c))},
\]
for any constant $1<cc_3\le c_0$.
Finally we deal with the $\log_2j$ terms in $A(x)$. 
Suppose there are $K_i$ datapoints stored in bucket $i$ at the end of the process.
Due to the Jensen's inequality,
\[
  \sum_{j=1}^{K_i}\log_2j\le K_i\log_2(K_i/2).
\]
Then
\[
  \Pr\left[\,\sum_i\sum_{j=1}^{K_i}\log_2j\ge cc_3\log u\,\right]\le \Pr\left[\,\sum_iK_i\log_2(K_i/2)\ge cc_3\log u\,\right].
\]
Observe that 
\begin{align*}
  &\Pr\left[\,K_i\log_2K_i\ge k\log_2k\,\right]\le\Pr\left[\,K_i\ge k\,\right]\le\binom{c_3\log u}{k}(1/\log u)^k\\
  \le&(ec_3/k)^k=2^{-k\log(k/ec_3)}.
\end{align*}
Note that $K_i$'s are negatively correlated, therefore 
\[
\Pr\left[\sum_iK_i\log_2K_i\ge cc_3\log u\right]\le\Pr\left[\sum_iK'_i\log_2K'_i\ge cc_3\log u\right], 
\]
where every $K'_i$ has the same distribution with $K_i$ but $K'_i$ are mutual independent.
Therefore
\[
  \Pr\left[\,\sum_iK'_i\log_2K'_i\ge cc_3\log u\,\right]\le\binom{(1+c)c_3\log u}{c_3\log u}u^{-(cc_3)+ec_3}\le u^{-(cc_3)+c_3\log(e(1+c))+ec_3}.
\]

Hence the sum of the lengths of the prefixes is upper bounded by $(c_3\log u)\lceil\log \log u\rceil+cc_3\log u$ with probability $1-u^{-cc_3/2}$ for large enough $c\le c_0$.
We then compress the headers (i.e. the $(c_3\log u)\lceil\log \log u\rceil$ term) with a bucket encoding:
We write down $0\circ 1^{K_0}\circ 0\circ 1^{K_1}\circ 0\circ \cdots 1^{K_{2^{\lceil\log \log u\rceil}-1}}\circ 0$ at first.
Then we enumerate bucket $i\in[2^{\lceil\log \log u\rceil}]$, and write down its local prefixes in lexicographical order with a special delimiter: $\perp\circ\alpha_i(X_{i_1})\circ\perp\circ\alpha_i(X_{i_2})\circ\perp\cdots\perp$.
To implement the special delimiter, we simply double the size of alphabet by denoting $0,1,\perp$ with $00,01,11$ respectively.

Finally, the space upper bound holds if $c_2>10 c c_3>30c_1$.

Fix $u, c_1, c_3$, the representation consists of two strings, the headers $hd$ and the bodies $bd$.
Note that $bd$ is an array of $\{0,1,\perp\}$.
Algorithm \ref{algorithm: adaptive prefixes lowerbound}, \ref{algorithm: adaptive prefixes lookup}, \ref{algorithm: adaptive prefixes insert} are the pseudocodes for lowerbound, lookup and insert algorithms of the prefix collection respectively.

\begin{algorithm}
  \SetKwProg{Fn}{Function}{ is}{end}
  \SetKw{Continue}{continue}
  \SetKw{Break}{break}

  \caption{The lowerbound query for adaptive prefixes}
  \label{algorithm: adaptive prefixes lowerbound}
  \SetKwInOut{Input}{input}\SetKwInOut{Output}{output}
  \SetKwData{True}{true}
  \Input{query $i$; headers $hd$;}
  \Begin
  {
	\Return $\mathtt{Select}(\neg hd,i+1)+1$\;
  }
\end{algorithm}

\begin{algorithm}
  \SetKwProg{Fn}{Function}{ is}{end}
  \SetKw{Continue}{continue}
  \SetKw{Break}{break}

  \caption{The lookup algorithm for adaptive prefixes}
  \label{algorithm: adaptive prefixes lookup}
  \SetKwInOut{Input}{input}\SetKwInOut{Output}{output}
  \SetKwData{True}{true}
  \SetKwData{YES}{YES}
  \SetKwData{NO}{NO}
  \Input{query $x$; headers $hd$, bodies $bd$;}
  \Begin
  {
	let $i,x_R$ be the left-most $\lceil\log\log u\rceil$ bits and the right-most bits of $x$ respectively\;
	$a\gets \mathtt{Select}(\neg hd,i+1),b\gets \mathtt{Select}(\neg hd,i+2)$\;
	$a'\gets\mathtt{Popcount}(hd[1,\ldots,a]), b'\gets\mathtt{Popcount}(hd[1,\ldots,b])$\;
	let $a'',b''$ be the indices of $(a'+1)$-th and $(b'+1)$-th $\perp$ in $bd$\;
	retrieve $bd[a'',\cdots,b''-1]$\;
	$j\gets\mathtt{Pred}(bd[a'',\cdots,b''-1],x_R)$\;
	\lIf{$j=0$}{\Return \NO}
	let $k,k'$ be the indices of $j$-th and $(j+1)$-th $\perp$ in $bd[a'',\cdots,b''-1]$ respectively\;
	$\alpha_y\gets bd[k+1,k'-1]$\;
	\lIf{$\alpha_y\not\sqsupseteq x$}{\Return \NO}
	\Return $a'+j$\;
  }
\end{algorithm}

\begin{algorithm}
  \SetKwProg{Fn}{Function}{ is}{end}
  \SetKw{Continue}{continue}
  \SetKw{Break}{break}

  \caption{The insertion algorithm for adaptive prefixes}
  \label{algorithm: adaptive prefixes insert}
  \SetKwInOut{Input}{input}\SetKwInOut{Output}{output}
  \SetKwData{True}{true}
  \Input{insertion $x$; headers $hd$, bodies $bd$;}
  \Begin
  {
	let $i,x_R$ be the left-most $\lceil\log\log u\rceil$ bits and the right-most bits of $x$ respectively\;
	$a\gets \mathtt{Select}(\neg hd,i+1),b\gets \mathtt{Select}(\neg hd,i+2)$\;
	$a'\gets\mathtt{Popcount}(hd[1,\ldots,a]), b'\gets\mathtt{Popcount}(hd[1,\ldots,b])$\;
	insert $1$ into $hd[b]$\;
	\lIf{$a=b$}{insert a $\perp$ into $bd[a']$ and \Return}
	let $a'',b''$ be the indices of $(a'+1)$-th and $(b'+1)$-th $\perp$ in $bd$\;
	retrieve $bd[a'',\cdots,b'']$\;
	$j\gets\mathtt{Pred}(bd[a'',\cdots,b''],x_R)$\;
	\If{$j\ne 0$}
	{
	let $k,k'$ be the indices of $j$-th and $(j+1)$-th $\perp$ in $bd[a'',\cdots,b'']$ respectively\;
	$\alpha_i(y)\gets bd[k+1,k'-1]$\;
	let $y'$ be the new adaptive prefix by extending $\alpha_i(y)$\;
	let $l,l'$ be the indices of $(j+1)$-th and $(j+2)$-th $\perp$ in $bd[a'',\cdots,b'']$ respectively\;
	\lIf{there exists $j+2$ $\perp$'s in $bd[a'',\cdots,b'']$}{ $\alpha_{y'}\gets bd[l+1,l'-1]$}
	\lElse{$\alpha_{y'}$ is empty string}
	compute $\alpha_i(x)$ with $\alpha_y$ and $\alpha_{y'}$\;
		\If{$\alpha_i(x)$ precedes $y'$}
	{
	  replace $bd[k+1,k'-1]$ with $y'$\;
	  insert $\perp\circ\alpha_i(x)$ into $bd[k+1]$\;
	}
	\Else
	{
	  insert $\perp\circ\alpha_i(x)$ into $bd[k']$\;
	  replace $bd[k+1,k'-1]$ with $y'$\;
	}
	}
	\Else
	{
	let $l'$ be the index of $2$-nd $\perp$ in $bd[a'',\cdots,b''-1]$\;
	$\alpha_{y'}\gets bd[a''+1,l'-1]$\;
	compute $\alpha_i(x)$ with $\alpha_{y'}$\;
	  insert $\perp\circ\alpha_i(x)$ into $bd[a'']$\;
	}

  }
\end{algorithm}

%
%
%
%
%
%
%
\section{Detailed Implementations and Pseudocodes}
\label{section: implementation}
We now describe the implementations of the subroutines.
\subsection{Initialize}
At first consider the works we should do to initialize a data structure $D$ with parameters capacity $m$, and datapoint length $\ell$.

We choose large enough $c_3,c_4,c_5$ to ensure the failure probability is at most $\delta$ according to Eq(\ref{eq: failure probability}), and compute the space usage for fingerprint collection according to Theorem \ref{theorem: adaptive-remainders}.
Then we create the main table with an extendable arrays of length $m/\log u$ with element size $w$.
In the next step, we initialize the subtables one by one.
For each subtable, we create an extendable array to maintain the fingerprints, the indicator list, the buffers, some local variables, and the pointers for navigators and data blocks.
The array for data blocks should be initialized as empty extendable array with element size $w$.

The way to implement this subroutine is clear now.
We set some flags as built-in variables to track the progress of the procedure, then do some work to carry the initialization on during each invoking of $\PMInit{\cdot}$.
It is easy to see that the procedure is finished within $O(m)$ times of invoking. 
\subsection{Destroy}
For every subtable, we should destroy the two extendable arrays for navigators and data blocks.
We then should destroy the extendable array used by the subtable itself.
Finally we should destroy the main table.
And the implementation is obvious.
The procedure is finished within $O(m)$ times of invoking. 
\subsection{Insert}
\label{section: implementation of insertion}
Recall that insertion algorithm is supposed to reorganize a dynamic data block in the background.
We introduce a built-in subroutine to support this.
The task list of the reorganization procedure is
\begin{enumerate}
  \item update $2$'s in the $I_i$ to $3$ and copy $B_{i,r}\gets B_{i,u}$ within single invoking.
  \item insert the index of the data block under reorganization into proper positions of the navigator list.
  \item sort the array of rest parts and the list of identities according to the pointers in the buffer $B_{i,r}$ while keep the pointers in the buffer updated.
  \item update $3$'s in the $I_i$ to $1$ within single invoking.
\end{enumerate}
All steps can be implemented easily, excepts the second one.
For the second step, we should insert $\log u/\log \log u$ values (at most $\log u$ bits in total) into the navigator list, so we extend the array of navigators at first.
In principle, the insertions should be done by scanning the navigator list and the indicator list in reversed order.
Without loss of generality, suppose we are inserting address $a$ and let $m'=|N_i|$ be the number of navigators.
Given a list of positions $(j_1,j_2,\cdots,j_{\log u/\log \log u})$, we set $j_{\log u/\log \log u+1}\gets m'+\log u/\log \log u+1$ be the end of the indicator list, set $j'\gets m'$ be the end of the navigator list.
Then we enumerate $k$ from $\log u/\log \log u$ to $1$:
given $k$, we move consecutive $N_i[j'-(j_{k+1}-j_k+1),\cdots,j']$ to $N_i[j_k+1,\cdots,j_{k+1}-1]$, set $N_i[j_k]\gets a$, then decrement $k$ by one and decrement $j'$ by $j_{k+1}-j_k+1$.
The procedure can be done in $\log u/\log \log u+c_3\log\log \log u$ time, since we either insert an address or move $\log u/\log\log \log u$ entries of $N_i$ with only one memory access. 

However it becomes more complicated if we want every subroutine to work in constant time.
In this case, insertions and queries may happen during the reorganization.
In other words, there may be $2$'s inserted into the indicator list, and queries may retrieve navigators during reorganization. 
To fix this issue, we works with three pointers $p, q, q'$ to track the progress.
The pseudocode for updating navigators is given in Algorithm \ref{algorithm: update navigators}.

\begin{algorithm}
  \SetKwProg{Fn}{Function}{ is}{end}
  \SetKw{Continue}{continue}
  \SetKw{Break}{break}

  \caption{Updating navigators}
  \label{algorithm: update navigators}
  \SetKwInOut{Input}{input}\SetKwInOut{Output}{output}
  \SetKwData{True}{true}
  \Input{ address $a$ }
  \Begin
  {
	$q\gets |N_i|+\log u/\log \log u$, $q'\gets |N_i|$; \tcp*[f]{$p$ has been initialized.}\\
	extend $N_i$ to leave space for new pointers\;
	set all items in $N_i[q'+1,\cdots,q]$ to $-1$\;
	\While{\True}
	{
	  let $p'$ be the index of the last $3$ in $I_i[1,\cdots,p-1]$\;
	  \lIf{no $3$ in $I_i[1,\cdots,p-1]$}
	  {
		\Break
	  }
	  let $l$ be the number of $1$'s in $I_i[p'+1,\cdots,p-1]$\;
	  \If{$l\le \log u/\log\log \log u$}
	  {
		copy $N_i[q'-l+1,\cdots,q']$ to a register $a'$\;
		set all items in $N_i[q'-l+1,\cdots,q']$ to $-1$\;
		copy $a'$ to $N_i[q-l+1,\cdots,q]$\;
		$N_i[q-l]\gets a$\;
		$p\gets p', q'\gets q'-l, q\gets q-l-1$\;
	  }
	  \Else(\tcp*[f]{too many pointers})
	  {
		$l\gets \log u/\log\log \log u$\;
		let $p''$ be the index of the $l$-th $1$ from the end in $I_i[p'+1,\cdots,p-1]$\;
		copy $N_i[q'-l+1,\cdots,q']$ to a register $a'$\;
		set all items in $N_i[q'-l+1,\cdots,q']$ to $-1$\;
		copy $a'$ to $N_i[q-l+1,\cdots,q]$\;
		$p\gets p''$, $q'\gets q'-l$, $q\gets q-l$\;
	  }
	}
	$p\gets 1$, $q\gets 1$\;
  }
\end{algorithm}

Therefore the built-in subroutine $\mathrm{reorganize}(i)$ can be designed similarly with previous subroutines so that the reorganization procedure can be done in $5\log u/\log \log u$ times of invoking.
%
The pseudocode for insertion is given in Algorithm \ref{algorithm: insert}.

\begin{algorithm}
  \SetKwProg{Fn}{Function}{ is}{end}

  \caption{Insertion algorithm}
  \label{algorithm: insert}
  \SetKwFunction{Retrieve}{Retrieve}
  \SetKwInOut{Input}{input}\SetKwInOut{Output}{output}
  \SetKwData{YES}{YES}
  \SetKwData{NO}{NO}
  \Input{insertion $x$}
  \Begin
  {
	$i\gets st(x)$\;
	$j\gets\mathrm{lookup}(\alpha_i,x)$\;
	\lIf{$j=$\NO}{ insert $hd(x)\circ hs(x)\circ rt(x)$ into fingerprint collection $\alpha_i$ }
	\Else
	{
	  $y\gets\Retrieve(j,i,hd(x),hs(x))$\;
	  insert $hd(x)\circ hs(x)\circ rt(x)$ into fingerprint collection $\alpha_i$ with $y$\; 
	  \tcc{$\alpha_i(y)$ should be extended}
	}
	let $j$ be rank of the new fingerprint\;
	insert $2$ into $I_i[j]$\;
	\lIf{$j\le p$}{$p\gets p+1$}
	\lIf{no data block under construction} { extend $A_i$ to create a new data block }
	\tcc{we leave space for headers and identities only}
	let $k$ be the address of data block under construction\;
	insert $1$ after the $(hd(x)+1)$-th $0$ to update the header list\;
	append $hs(x)$ and $rt(x)$ to the corresponding lists of data block $k$\;
	let $k'$ be index of $rt(x)$\;
	let $j'$ be the number of $2$'s in $I_i[1,\ldots,j]$\;
	insert $k'$ into $B_{i,u}[j']$\;
	\If{data block $k$ is full}
	{
	  update the $2$'s in the $I_i$ to $3$\;
	  $B_{i,r}\gets B_{i,u}$\;
	  $p\gets |I_i|+1$\;
	}
	invoke $\mathrm{reorganize}(i)$ for $10$ times\;
  }
\end{algorithm}

\subsection{Query}
\label{section: implementation of query}
The query algorithm is formally described in Algorithm \ref{algorithm: query}.

\begin{algorithm}
  \SetKwProg{Fn}{Function}{ is}{end}

  \caption{Retrieve algorithm}
  \label{algorithm: retrieve}
  \SetKwFunction{FindNavigator}{FindNavigator}
  \SetKwFunction{Retrieve}{Retrieve}
  \SetKwInOut{Input}{input}\SetKwInOut{Output}{output}
  \SetKwData{YES}{YES}
  \SetKwData{NO}{NO}
  \Output{$y$ which is corresponding to the $j$-th fingerprint}
  \Fn{\Retrieve{fingerprint rank $j$; subtable index $i$; header and identity $hd,hs$}}
  {
	\If{$I_i[j]=1$}
	{
	  $j'\gets$ \FindNavigator{$j$}\;
	  $k\gets N_i[j']$\;
	  $a\gets$ \FindNavigator{$\mathrm{lowerbound}(\alpha_i,hd\circ hs)$}\;
	  let $k'$ be the number of $k$'s in $N_i[a,\cdots,j']$\;
	  let $h$ be the binary representation of header list of data block $k$\;
	  let $a_1,a_2$ be the indices of $(hd+1)$-th and $(hd+2)$-th $0$ in $h$ respectively\;
	  let $b_1,b_2$ be the numbers of $1$ in $h[1,\cdots,a_1]$ and $h[1,\cdots,a_2]$ respectively\;
	  let $s$ be the identity list of data block $k$\;
	  compute $k''$ such that $s[k'']$ is the $k'$-th $hs$ in $s[b_1+1,\cdots,b_2]$\;
	  let $rt_{k''}$ be the $k''$-th rest part\;
	  \Return $i\circ hd\circ hs\circ rt_{k''}$\;
	}
	\Else
	{
	  \If{$I_i[j]=2$}
	  {
		let $k$ be the address of data block under construction\;
		let $j'$ be the number of $2$'s in $I_i[1,\cdots,j]$\;
		$k'\gets B_{i,u}[j']$\;
	  }
	  \Else
	  {
		let $k$ be the address of data block under reorganization\;
		let $j'$ be the number of $3$'s in $I_i[1,\cdots,j]$\;
		$k'\gets B_{i,r}[j']$\;
	  }
	  \lIf{data block $k$ or $k'$-th rest part in data block $k$ is deleted}{\Return \NO}
	  let $rt_{k'}$ be $k'$-th rest part in data block $k$\;
	  \Return $i\circ hd\circ hs\circ rt_{k'}$\;
	}
  }
  \Fn{\FindNavigator{prefix rank $j$}}
  {
	let $j',j''$ be the number of $1$'s and $3$'s in $I_i[1,\cdots,j]$ respectively\;
	\lIf{$p>j$}
	{\Return $N_i[j']$}
	\lElse
	{\Return $N_i[j'+j'']$}
  }
\end{algorithm}
\begin{algorithm}
  \SetKwProg{Fn}{Function}{ is}{end}

  \caption{Query algorithm}
  \label{algorithm: query}
  \SetKwFunction{FindNavigator}{FindNavigator}
  \SetKwFunction{Retrieve}{Retrieve}
  \SetKwInOut{Input}{input}\SetKwInOut{Output}{output}
  \SetKwData{YES}{YES}
  \SetKwData{NO}{NO}
  \Input{query $x$ }
  \Output{whether there is a prefix of $x$ in database}
  \Begin
  {
	$i\gets st(x)$\;
	$j\gets\mathrm{lookup}(\alpha_i,x)$\;
	\lIf{$j=$\NO}{ \Return \NO}
	$y\gets\Retrieve(j,i,hd(x),hs(x))$\;
	\lIf{$rt(y) \sqsupseteq rt(x)$} { \Return \YES}
	\lElse{\Return \NO}
  }
  \end{algorithm}
\subsection{Decrement}
We maintain a $i$ that traverses from $1$ to $m/\log u$.
For a subtable which has $m'$ data blocks, 
we maintain a $j$ that traverses $m'$ to $1$.
For a data block which has $m''$ datapoints, we maintain a $k$ that traverses from $m''$ to $1$. 

For a dynamic data block $j$, we scan the corresponding buffer to look for a $j'$ such that $j'$-th entry is $k$,
scan the indicator list to look for a $j$ such that $I_i[j]$ is the $j'$-th $2$ (or $3$, depends on the type of the dynamic data block), retrieve the $hd(y)$ from $j$-th fingerprint, then output $i\circ hd(y)\circ hs_k\circ rt_k$ and shrink the extendable array to free the space occupied by $rt_k$.
And for a static data block $j$, we output $i\circ hd_k\circ hs_k\circ rt_k$, where $hd_k, hs_k,rt_k$ are the $k$-th header, identity and rest part respectively, then shrink the extendable array to free the space occupied by $rt_k$.

\end{document}